\documentclass{ws-procs975x65-ar}

\def\Journal#1#2#3#4{{#1} {\bf #2}, #3 (#4). }

\def\PLB{{\em Phys. Lett.} B}

\def\PRL{\em Phys. Rev. Lett.}

\def\PRD{{\em Phys. Rev.} D}

\def\JINST{{\em JINST}}

\begin{document}

\title{NEUTRINOS OSCILLATIONS WITH LONG-BASE-LINE BEAMS\\
(Past, Present and very near Future)\footnote{To be published in the Proceedings of the
Int. Conf. Beyond the Standard Models of Particle Physics, Cosmology and Astrophysics (BEYOND 2010), 1-6 February 2010, Cape Town, 
South Africa.}}

\author{L.STANCO}

\address{INFN - Padova, www.pd.infn.it\\
Via Marzolo, 8, Padova I-35131 Italy\\
E-mail: luca.stanco@pd.infn.it
}

\begin{abstract}
We overview the status of the studies on neutrino oscillations with
accelerators at the present running experiments. Past and present results
enlighten the path towards the observation of massive neutrinos and 
the settling of their oscillations. The very near future may still have 
addiction from the outcome of the on-going experiments. OPERA is chosen as
a relevant example justified by the very recent results released.
\end{abstract}


\bodymatter

\section{Introduction}\label{aba:sec1}

In the last two decades several experiments have provided strong evidence in favor of the
neutrinos oscillation hypothesis. In the so called {\em atmospheric sector} the flavor conversion 
was first established by Super-Kamiokande\cite{sk} and further by MACRO\cite{macro} and Soudan-2\cite{soudan} experiments.
Further confirmation was more recently obtained by the K2K\cite{k2k} and MINOS\cite{minos} 
long-baseline experiments. However a two fold question is still unanswered, does the oscillation scenario 
correspond to the simple 3-flavor expectation or not? which is related to the still unobserved 
direct appearance of one flavor to another, in particular to the highly expected $\nu_\mu\rightarrow\nu_\tau$
oscillation. Answer to this two-fold question is relevant mainly to proceed towards the next steps in the
clarification of the leptonic sector of the particle model.

After a brief reminder of the physics behind we will assay to focus on the main points which brings us to the 
present knowledge about  neutrino mixing. The recent history provided the scenario in which the neutrino
oscillation framework was settled. Still new questions opened up and these bring us directly into the future.
Next we will shortly report on the present results from short-base-line (SBL) experiment, mainly the 
MiniBooNE\cite{miniboone}
experiment, and the long-base-line (LBL) experiments, namely MINOS and OPERA\cite{opera1}.
Finally some physics expectations for the near future after a personal discussion of the very recent
OPERA results\cite{opera3} will be drawn.

\section{Physics layout}\label{aba:sec2}

The issue of the lepton mixing is far from being understood and even generally described as
it occurs in the quark sector. In particular the generic three questions on the reason
the leptons mix themselves, the details of the way they actually mix and which are the mechanisms
which underlay their mixing, arize.
In 1998 a new history for neutrinos began as a sort of second life with the double discovery that
(a) they oscillate\cite{sk} then owing a mass after 41 years from the initial idea of B. Pontecorvo in 1957\cite{pontecorvo}
and (b) they mix themselves in a peculiar way after the void result by CHOOZ\cite{chooz}. 

The CHOOZ experiment took data in 1997-98 at a distance of about 1 km from a nuclear power plant of two reactors in France.
It aimed to observe $\nu_e \rightarrow \nu_{\mu}$ (actually antineutrinos) oscillations. After a collection
of 2991 ${\bar \nu_e}$ candidates CHOOZ put an upper limit on the direct observation of  ${\bar \nu_{\mu}}$ events.
At that time the limit was set as $\sin\theta \lesssim 0.1$ with a systematic error of 2.7\%. The low error was due to the possibility for 
CHOOZ to measure the backgrounds before the switching on of the reactors.

In 2002 the KamLAND experiment\cite{kamland} repeated the measure in a site in Japan where many reactors were present,
close and far away from the detector. The distribution of the ${\bar \nu_e}$  flux coming from the reactors 
is displayed in Fig.~\ref{fig1.2}(a), with an average distance of 150 km from the reactor. Differently from CHOOZ, KamLAND
obtained a positive result in term of disappearance of ${\bar \nu_e}$  flux. The beautiful oscillation pattern is shown
in Fig.~\ref{fig1.2}(b).

\def\figsubcap#1{\par\noindent\centering\footnotesize(#1)}
\begin{figure}[htb]%
\begin{center}
  \parbox{2.1in}{\epsfig{figure=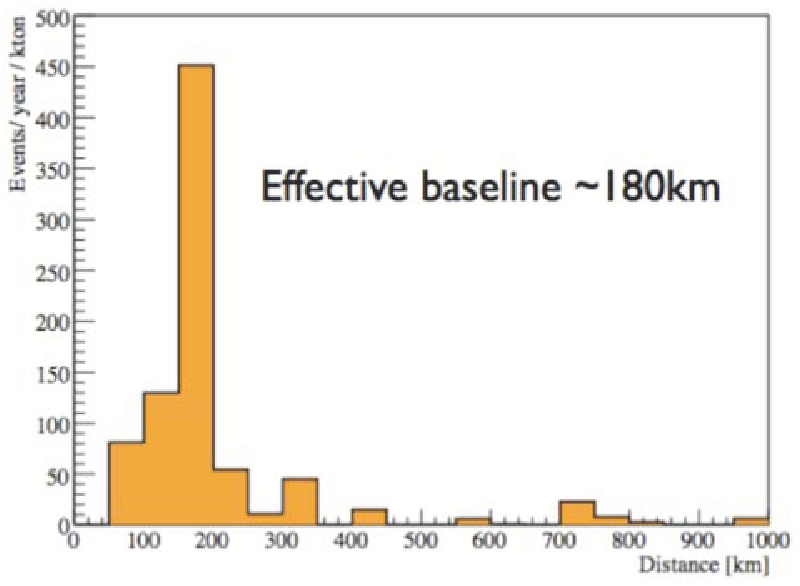,width=2.2in}\figsubcap{a}}
  \hspace*{4pt}
  \parbox{2.1in}{\epsfig{figure=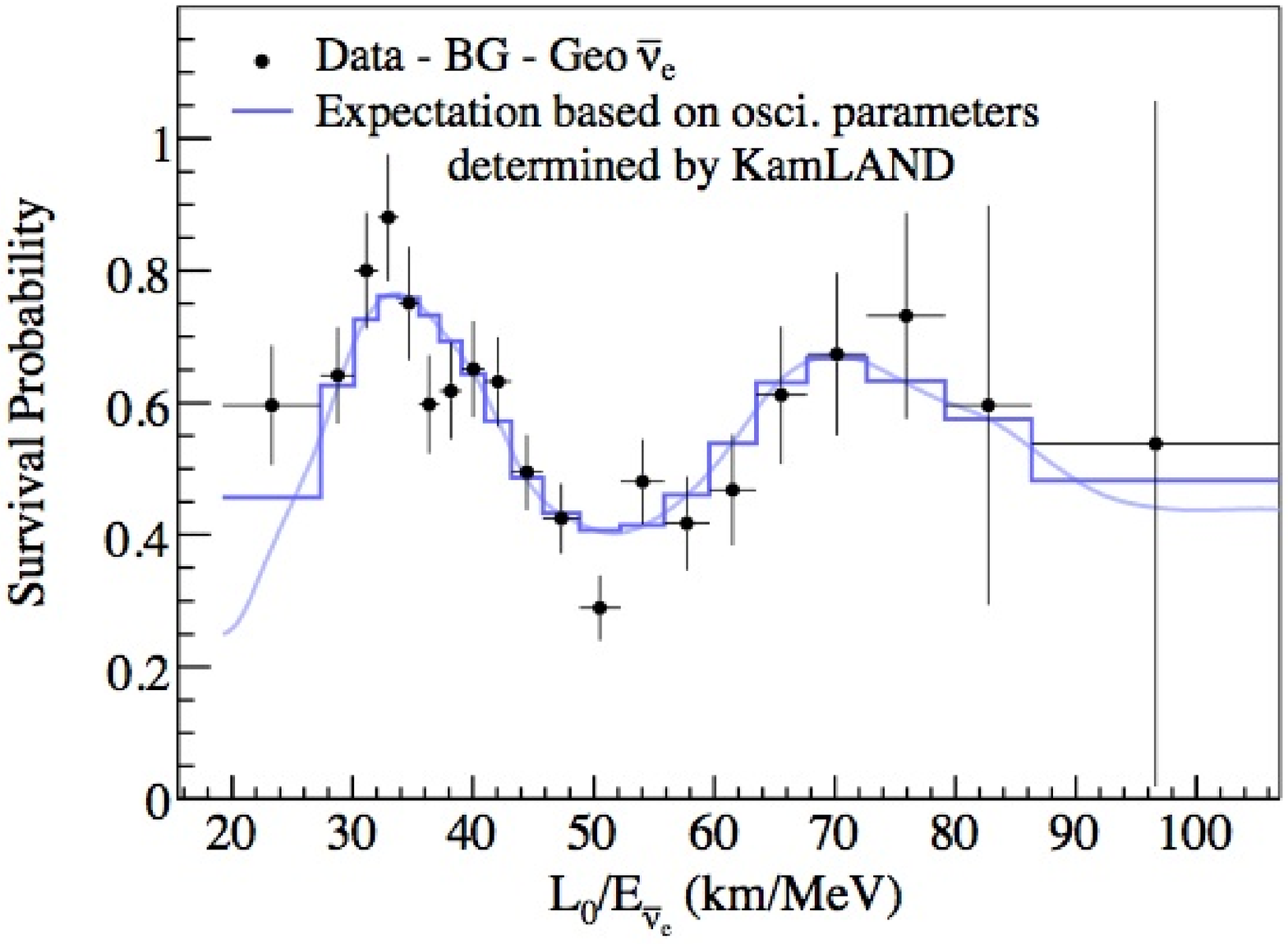,width=2.5in}\figsubcap{b}}
  \caption{The KamLAND experiment. (a) Distribution of the ${\bar \nu_e}$  flux. (b) Oscillation pattern of the ${\bar \nu_e}$ 
  disappearance. The figures are taken from [\refcite{kamland}].}%
  \label{fig1.2}
\end{center}
\end{figure}

Mainly after KamLAND (and a rather contemporary result in the {\em solar} neutrino sector by the SNO experiment\cite{sno})
the increase in the oscillation neutrino studies was extremely rapid and huge
bringing to a re-interpretation of the CHOOZ result in term of oscillations of flavour eigenstates.
The old idea of mixing matrix by Maki et al. in 1962\cite{mns} was  revitalized, similarly to what was made by 
Cabibbo\cite{cabibbo} in 1963 for the quark sector. The standard parametrization of a mixing matrix at 3
components is therefore realized via the usual 3 Euler rotations, leaving us with 3 angles,
$\theta_{12}$, $\theta_{23}$, $\theta_{13}$, and a phase $\delta$. Moreover in case of a Majorana picture 
two more phases are present, $\alpha_1$ and $\alpha_2$. To emphasize the key point it comes out quite
naturally to simply establish a similar way of mixing for
quarks and leptons. Of course other more complex scenarios, where more than 3 eigenstates appear,
are possible. More neutrinos states are compatible with the present knowledge of the lepton physics,
in particular one or more {\em sterile} neutrinos\cite{sterile} may be included. This is a fundament question
since it may or it may not en strength parallelism between quarks and leptons.
 
The complete description of the formalism may be found in [\refcite{strumia}], while several fits have been performed
to take into account the whole set of measurements. Still fundamental questions remain unanswered.
The first question relates to the mass ordering of the neutrino mass eigenstates. Does the mass scale ordering of
$\nu_1$, $\nu_2$, $\nu_3$ (as defined by the parametrization) follows the same ordering of $\nu_e$, $\nu_{\mu}$, 
$\nu_{\tau}$ ? As the measured oscillation pattern is described only in term of $\Delta m^2_{12}$ and $\Delta m^2_{13}$
the exact order is not identified yet, neither it is the absolute mass scale. Are the 3 masses just below the
present neutrino mass absolute limit (less than 1 eV) or are they some order of magnitude smaller?

More and more unanswered questions come up as we put a closer look to the measured quantities.
For example in Table~\ref{tab1} the present values of the mixing matrix components for quarks, $V_{CKM}$, and
for leptons, $V_{MNS}$ are compared. The underlying pattern is clearly different and we finally conclude that
the lepton mixing is weird\footnote{Even if the lepton mixing appears weird several tentatives to elaborate
a quark-lepton complementarity by playing on the relative values of the $\theta$'s angles have been done.
See for example Ref.~\refcite{ferruccio}.}.

\begin{table}
\tbl{Present values for the Neutrino Mass Mixing Matrix (a) as taken from Ref.~\refcite{mezzetto} 
and the unitarity values of the $V_{CKM}$ (b) as extracted from Ref.~\refcite{PDG}. Note that the very recent result by
MINOS\cite{minos-nue1} sets $\sin^2 2\theta_{13} < 0.12$.} 
{\begin{tabular}{@{}cccc@{}}
\toprule
(a) & $ \sin^2 \theta_{12}$ & $=$ & $0.30\pm 0.02 $ \\
& $\sin^2 \theta_{23}$ & $=$ & $0.50\pm 0.07 $ \\
& $\sin^2 2\theta_{13} $ & $ < $ & $0.13$ \\
& $\Delta m^2_{13}$ & $=$ & $2.40^{+0.12}_{-0.11} \; \times 10^{-3}\, eV^2$ \\
& $\delta m^2_{12}$ & $=$ & $7.6\pm 0.2 \; \times 10^{-5}\, eV^2$ \\
%\colrule
\botrule
(b) & $\sum_{i=d,s,b} |V_{ui}|^2$ & $=$ & $0.9999\pm 0.0011 $ \\
 & $\sum_{i=u,c,t} |V_{id}|^2$ & $=$ & $1.002\pm 0.005$ \\
 & $\sum_{i=u,c; j=d,s,b} |V_{ij}|^2$ & $ =$ & $2.002\pm 0.027$ \\
\botrule
\end{tabular}
}
\label{tab1}
\end{table}

\begin{figure}[htb]
\begin{center}
\psfig{file=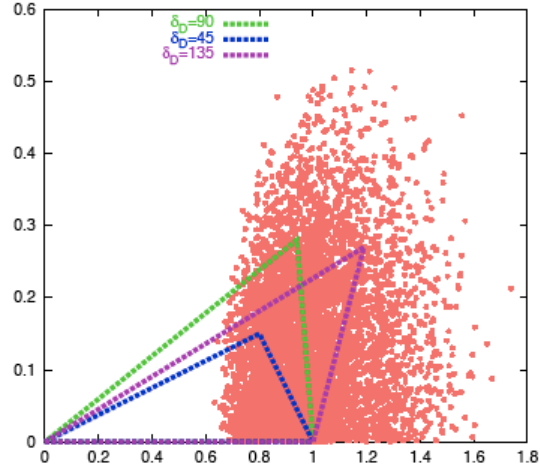,width=3in}
\end{center}
\caption{The unitarity $e\mu$-triangles. The horizontal side, $|U_{e1}U_{\mu1}^*|$ is normalised to one. The triangles correspond 
to $\theta_{13} = 0.15$ and different values of the phase $\delta$. Each scatter point represents a possible position of vertex as
the mixing parameters 
pick up random values within the present uncertainty ranges: $\sin^2\theta_{23} \in [0.36, 0.61]$, 
$\sin^2\theta_{12} \in [0.27, 0.37]$ and $\sin^2\theta_{13} \in [0, 0.031]$, and $\delta \in [0, 2\pi]$. There are also illustrated
3 different triangles for 3 different choices of $\delta$ and $\theta_{13}=8.6^0$ case. The figure is taken from [\refcite{ISS}].}
\label{figISS}
\end{figure}

Also the present knowledge of the errors is largely different in the quark and lepton sector. See e.g.  Ref.~\refcite{mezzetto}
for an up-to-date report on the error measurements, to be compared with the extremely well known values
of the quark mixing matrix\cite{PDG}. To illustrate the importance of the size of the errors we may look at Fig.~\ref{figISS}
taken from Ref.~\refcite{ISS} (Fig.~43), which shows the large region for the possible values of the top angle of the lepton unitarity
triangle. The degenerate case, $\theta_{13} = 0$, corresponding to the bottom horizontal line, is also still allowed 
by the present measurements.
More and exhaustive discussions may be found in Ref.~\refcite{farzan}.

In summary we may conclude that the lepton mass mixing matrix might be technically similar to the quark one even if 
it shows a quite different pattern and it is at present rather poor known. We like to conclude this section by 
using the same wording of W. Buchm\"{u}ller at EPS09 conference\cite{EPS}: "Right-handed neutrinos have been found;
no exotics have been found (yet)". Therefore as a whole it follows that we have to be prepared to the unexpected!

\section{Physics perspectives}

Currently the lepton scenario illustrated in the previous section is the only one which is receiving attention by
experimental investigation and mostly phenomenological investigation too. Other theoretical possibilities like e.g. 
the NSI, Non-Standard-Interactions\cite{NSI}, are in our judgement not so appealing and remains at the level of 
generic phenomenological models.

Therefore a not so long list of unknowns have to be identified and measured: the 3 mixing angles ($\theta_{12}$, $\theta_{13}$ 
and $\theta_{23}$), 
the 2 neutrino squared mass differences ($\Delta m^2_{12}$, $\Delta m^2_{13}$), the sign of one the two mass differences
($\Delta m^2_{23}$), a CP phase ($\delta$),
the absolute neutrino mass scale and their nature (Dirac or Majorana), the total number of neutrino (are there
more than 3 neutrinos ? \footnote{The possibility of more than 3 neutrinos refers to the presence of the so called 
 {\em sterile} neutrinos\cite{sterile}, id est neutrinos not active from the point of view of the weak interaction.}), 
 not at last forgetting the detection of the undergoing source of the oscillation. 
The latter question corresponds to the detection of a direct appearance signal, that is the observation of the $\nu_{\tau}$
appearance for the atmospheric oscillation (and the $\nu_e$ for the solar one) providing a direct measurement of the Lepton Flavor Violation (LFV)
process\footnote{The SNO experiment\cite{sno} measured the appearance of neutrinos with flavor different from the original
electronic one in the {\em solar} sector. We name this kind of observation {\em indirect}.}.

\noindent Most of the above items may be investigated at Long-Base-Line experiments by excluding the investigation of 
the fundamental nature of the neutrinos and their absolute mass scale.

\begin{figure}[htb]
\begin{center}
\psfig{file=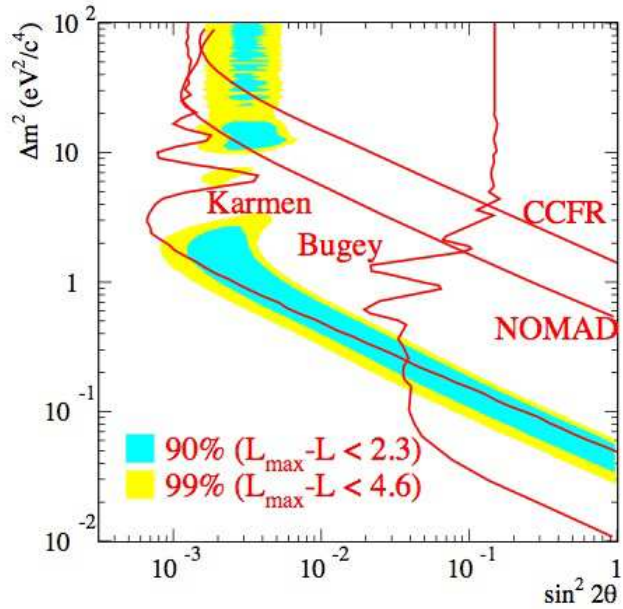,width=3.5in}
\end{center}
\caption{The LSND observation limits of the ${\bar \nu_{\mu}}-{\bar \nu_e}$ oscillation. 
The allowed regions are obtained by a $(\sin^2 2\theta, \Delta m^2)$ oscillation parameter fit, at 90\% and 95\% C.L.
The curves are 90\% CL limits from the Bugey reactor experiment, the CCFR experiment at Fermilab, 
the NOMAD experiment at CERN, and the KARMEN experiment at ISIS.
The figure is taken from [\refcite{lsnd}].}
\label{figLSND}
\end{figure}

\begin{figure}[htb]%
\begin{center}
  \parbox{2.1in}{\epsfig{figure=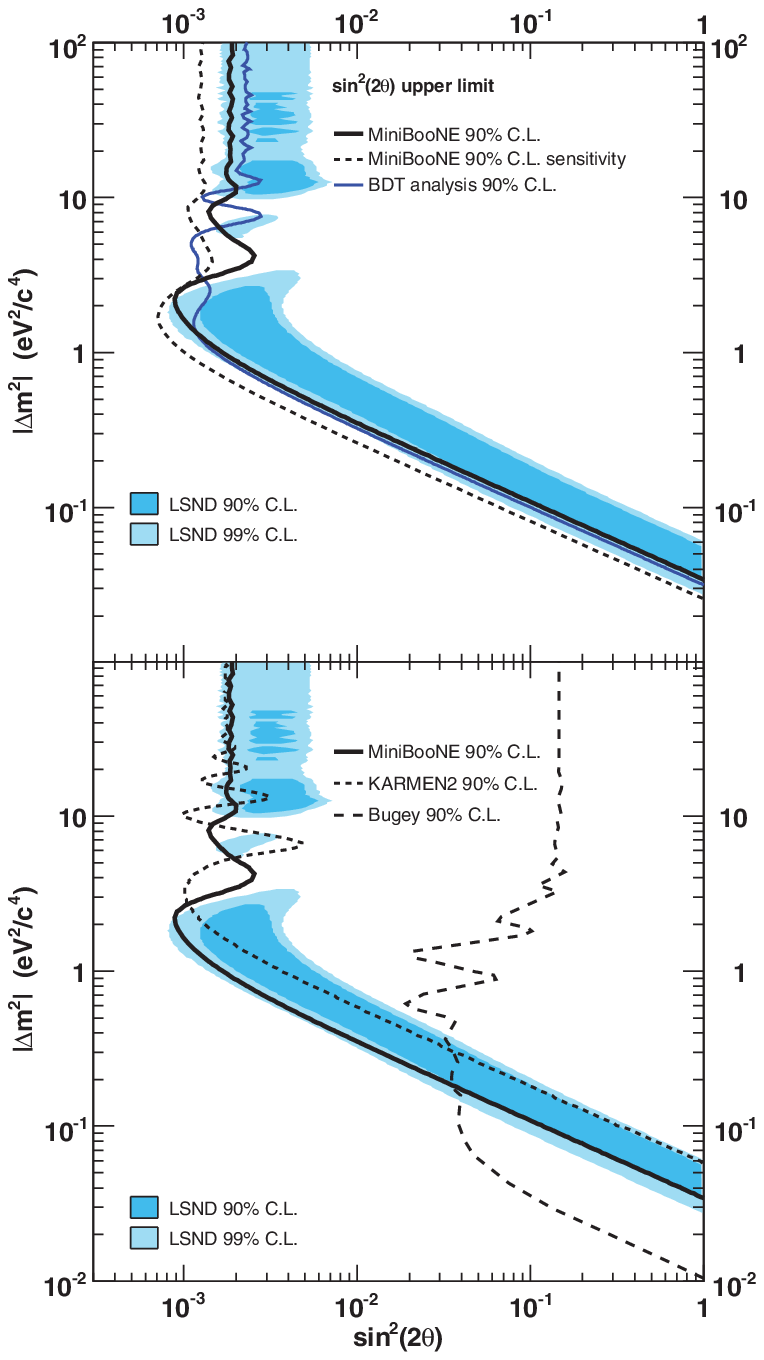,width=2.4in}\figsubcap{a}}
  \hspace*{4pt}
  \parbox{2.1in}{\epsfig{figure=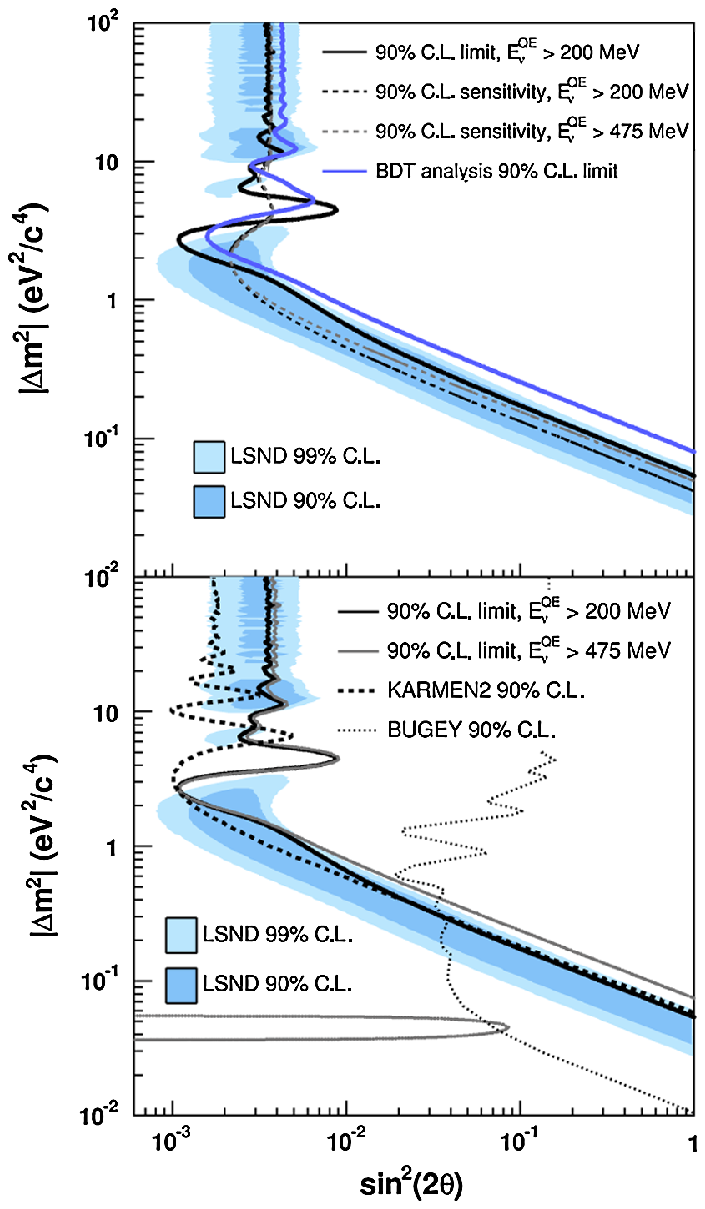,width=2.5in}\figsubcap{b}}
  \caption{The MiniBooNE experiment. (a) The limits extracted from the neutrino data ($5.58\pm 0.12)\times 10^{20}$ proton-on-target (p.o.t.)
   (b) The limits extracted from the antineutrino data ($3.39\times 10^{20}$ p.o.t.). The figures are taken from [\refcite{miniboone}].}
  \label{figMiniBooNE}
\end{center}
\end{figure}

The physics prospects are raveled by the "presence" of internal puzzles in the experimental side. In particular the
recent results from MiniBooNE are not able to disentangle the somewhat old and controversial result by LSND\cite{lsnd}.
The original result from LSND (see Fig.~\ref{figLSND}) of the ${\bar \nu_{\mu}}-{\bar \nu_e}$ observation
could not be phenomenologically arranged in the 3 neutrino standard scenario. MiniBooNE\cite{mini-app} looked for the oscillation
in either the neutrino or the antineutrino modes. In the neutrino mode it is able to rule out the result by LSND as oscillation
while observing an unexplained excess in a energy region below that of LSND. In the antineutrino mode no similar
excess is observed while the ruling out of LSND is not gained.  Fig.~\ref{figMiniBooNE} (a and b) as extracted by
Ref.~\refcite{miniboone} shows the MiniBooNE results.

As a matter of fact to the author the experimental situation is rather confused. More experimental facts are needed and the question
whether the ongoing two LBL experiments MINOS and OPERA may help turns out to be fully relevant.

\section{MINOS physics results}

\begin{figure}[htb]%
\begin{center}
  \parbox{2.1in}{\epsfig{figure=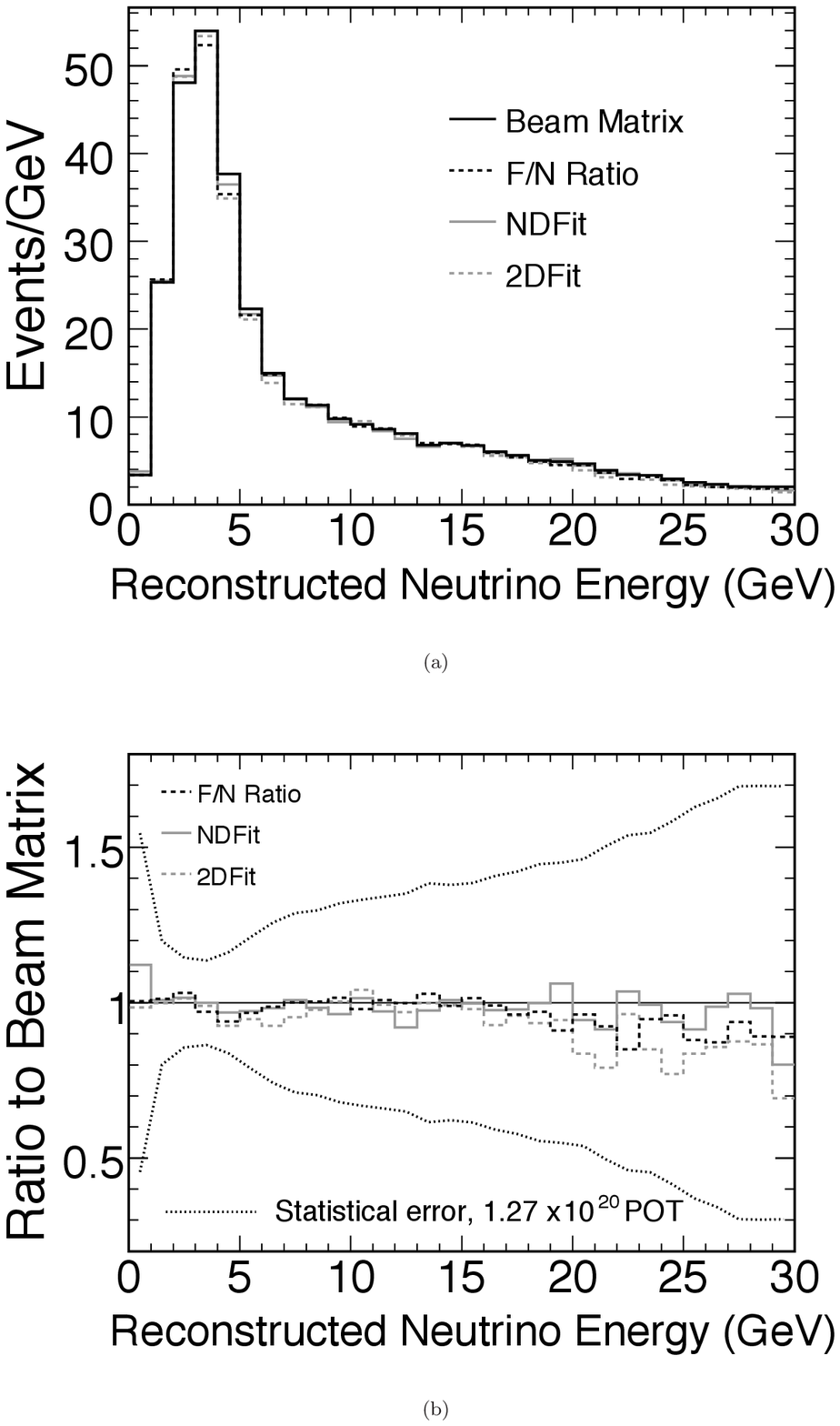,width=2.4in}\figsubcap{A}}
  \hspace*{4pt}
  \parbox{2.1in}{\epsfig{figure=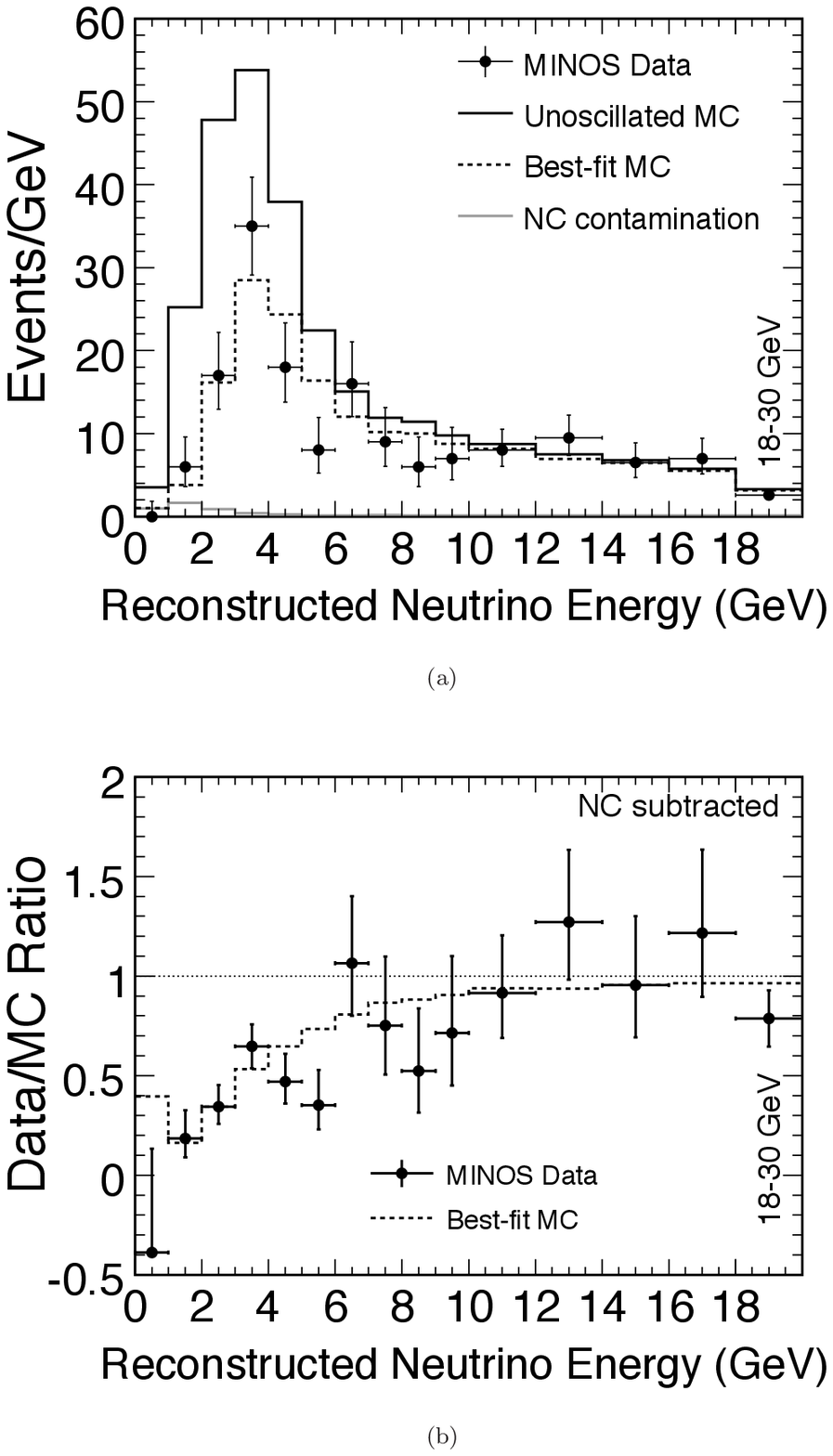,width=2.4in}\figsubcap{B}}
  \caption{The MINOS experiment. (A) Neutrino energy spectra at the Far Detector in the absence of neutrino oscillations 
  as predicted by the four extrapolation methods used.
  The limits are extracted from the neutrino data ($5.58\pm 0.12)\times 10^{20}$ proton-on-target (p.o.t.)
   (B) The reconstructed energy spectra of selected Far Detector events with the Far Detector unoscillated prediction (solid histogram) and best-fit 
   oscillated spectrum (dashed histogram) overlaid and (b) the ratio of the observed spectrum to the unoscillated Far Detector prediction.
   The figures are taken from [\refcite{minos1}].}%
  \label{figMinos}
\end{center}
\end{figure}

\begin{figure}[htb]
\begin{center}
\psfig{file=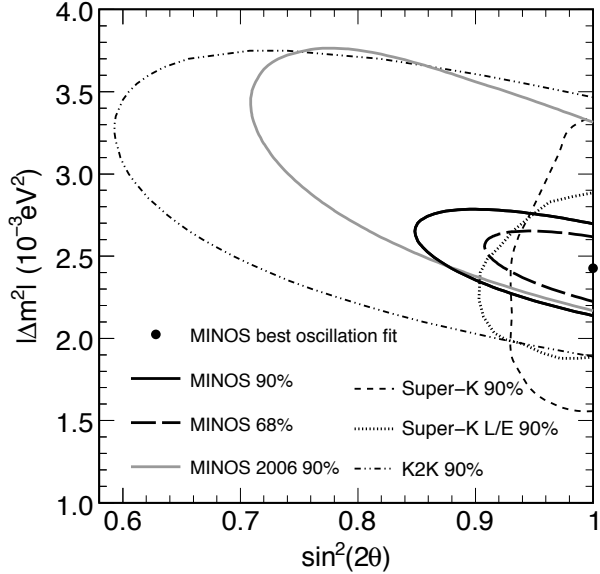,width=3.5in}
\end{center}
\caption{The MINOS observation limits of the ${\nu_{\mu}}$ oscillation.  Contours for the oscillation fit to the data in Fig. \ref{figMinos}-B. 
Also shown are contours from Super-K and K2K and earlier MINOS result in 2006.
The figure is taken from [\refcite{minos}].}
\label{figMinos1}
\end{figure}

The MINOS experiment\cite{minos-app} is constituted by two similar apparata, the Near and the Far detectors, made of 
scintillator strips and a toroidal spectrometer. This layout allows the minimization of several uncertainties like 
the neutrino flux from the NUMI beam and the extrapolation via Monte Carlo of the unoscillated $\nu_{\mu}$ spectrum from 
Near to Far sites. 
A very detailed analysis allows to reconstruct the energy of the interacting neutrinos (Fig.~\ref{figMinos}) and
estimate the percentage of disappeared neutrinos\cite{minos1}. From the later MINOS extracts the oscillation 
parameters in the assumption of 2 flavor oscillation mode (Fig.~\ref{figMinos1} from the analysis in [\refcite{minos}]).

\begin{figure}[htb]
\begin{center}
\psfig{file=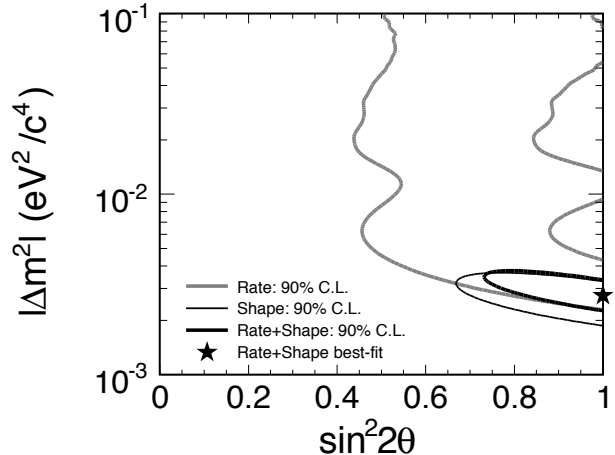,width=3.5in}
\end{center}
\caption{MINOS result: comparison of the 90\% C.L. regions from oscillation fits using shape and rate information, apart and together. The
best-fit point and 90\% C.L. contour from the fit to shape and rate information are also shown.
The figure is taken from [\refcite{minos1}].}
\label{figMinos2}
\end{figure}

Since we will discuss in the next section the OPERA experiment it is worthwhile to outline the twofold character of the
MINOS analysis, the "rate" and the "shape". As OPERA will be able to deal only with "rates", the latter significance power
has to be compared
with the corresponding one by MINOS which turned out to be rather poor (Fig.~\ref{figMinos2}).

The disappearance mode can be complementary studied in MINOS with the appearance of electron $\nu$.
First results reported were indicative of a possible $\nu_e$ appearance: 35 events from $\nu_e$ interactions were observed against 
an expected background of $27\pm 5(stat) \pm 2(sys)$, corresponding to a 1.5 excess\cite{minos-nue}. However very recent results (released
after the Conference time) with an increased statistics washed out that indication\cite{minos-nue1}. It seems that the new dedicated
experiments for the $\theta_{13}$ measurement have to be waited for (see the related contributions to these proceedings).

\section{The OPERA way}

We will now discuss at length the OPERA experiment since the very recent on May 31$^{rst}$ 2010 release of new results (see next Section)
corresponds to a relevant new contribution in the neutrino physics. 

The OPERA experiment\cite{opera1} has been designed to observe the
$\nu_\tau$ appearance in the CNGS $\nu_\mu$ beam\cite{cngs} on an event by event basis. 
The $\nu_\tau$ signature is given by the decay topology of the short-lived $\tau$ leptons produced in the $\nu_\tau$ Charged Current (CC) interactions decaying to one prong (electron, muon or hadron) or three prongs hadrons.
The detector is located underground in the Laboratorio Nazionale del Gran Sasso (LNGS, L'Aquila, Italy) along the path of the 
CNGS neutrino beam, 730 km away from the source at CERN. 
The beam was optimized in order to maximize the number of $\nu_\tau$ CC interactions at the LNGS site keeping the energy constraint to 
be above the $\tau$ production threshold. The result is a wide band neutrino beam with an average energy of $\sim 17$
 GeV; the $\bar{\nu}_\mu$ contamination is 2.1\%,  $\nu_e +\bar{\nu}_e$ is below 1\% and prompt $\nu_\tau$ at production is negligible. 
 With a nominal beam intensity of $4.5\times 10^{19}$ proton-on-target (p.o.t.) per year, 
 $\nu_\mu$ CC and neutral current (NC) interactions at Gran Sasso are deemed to ~2900/(kton$\times$year) and 875/(kton$\times$year), 
 respectively. By
 assuming the oscillation parameters $\Delta m^2 = 2.5\times 10^{-3}$ eV$^2$ at full mixing 10.4 events are expected to be observed in OPERA in 5 years of data 
 taking with a background of  0.75 events. 

\begin{figure}[hbt]
\begin{center}
\psfig{file=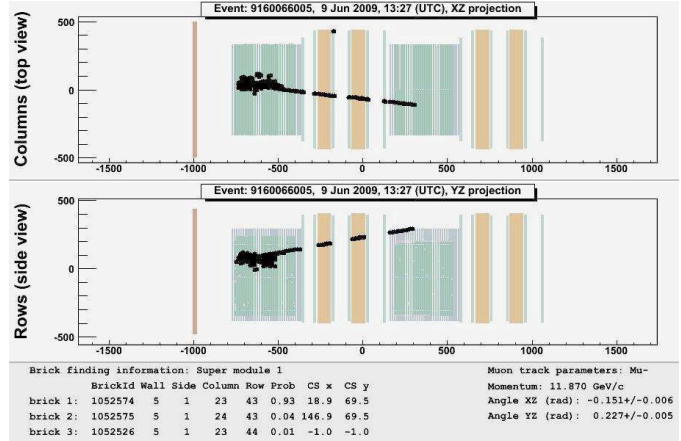,width=3.5in}
\end{center}
\caption{Neutrino event from OPERA as registered by the electronic detectors. The figure is taken from  [\refcite{luca}].}
\label{f2}
\end{figure}

In the two years 2008 and 2009 OPERA succeeded\cite{luca} to collect $5.30\times 10^{19}$ p.o.t. corresponding to 31,550 detected events 
in time with the 
beam, 5391 of which matched to a neutrino interaction in the OPERA target within more than 99\% percent accuracy.
At the CNGS energies the average $\tau$ decay length is $\sim 450~\mu\mbox{m}$. 
In order to observe it OPERA makes use of $2\times 44 \mu m$ nuclear emulsions films
interspaced with $1~\mbox{mm}$ thick lead plates which form the target mass of the OPERA detector. This technique, 
called Emulsion Cloud Chamber (ECC), has been used successfully by the DONUT experiment for the first direct observation\cite{DONUT} 
of the $\nu_\tau$.
Every time a trigger in the electronic detectors is compatible with an interaction inside the target (see Fig.~\ref{f2}), the brick with the highest 
probability to contain the
 neutrino interaction vertex is extracted from the apparatus and exposed to X-rays for film-to-film alignment. 
 Further the brick is unsandwiched, the emulsion films are developed and analyzed. The final sensitivities are
 $\sim$0.3 $\mu$m spatial resolution, $\sim$2 mrad angular resolution and $\sim$90\% single track detection efficiency.
 
\begin{figure}[htb]
\begin{center}
\psfig{file=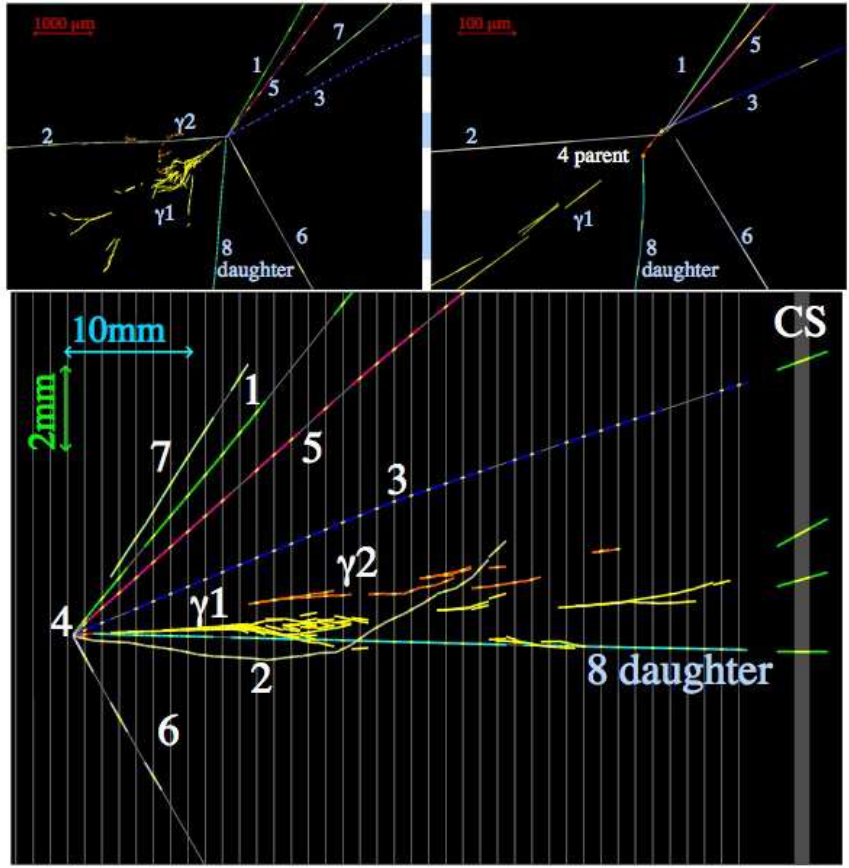,width=4.5in}
\end{center}
\caption{Display of the $\nu_{\tau}$ candidate event. Top left: view transverse to the neutrino direction. 
Top right: same view zoomed on the vertices. Bottom: longitudinal view. The figure is taken from  [\refcite{opera3}].}
\label{tauevt}
\end{figure}
\vskip 15pt

\noindent{\bf ADDENDUM}
Very recently OPERA reported the observation of a first $\nu_{\tau}$ candidate\cite{opera3}.  The result is obtained by the observation of a rather clean
event (Fig.~\ref{tauevt}), a possible 1-prong hadron decay of a $\tau$ lepton with $(n)\pi_0$ derived by the presence of some electromagnetic 
showers. The decay topology is consistent 
to be that of $\tau^-\rightarrow\nu_{\tau}+\rho^-\rightarrow\nu_{\tau}+\pi^-\pi_0$. Even if the expected number of $\nu_{\tau}$ interactions
and identification in OPERA is estimated
to be $0.54\pm 0.13$, well in agreement with the possible observation of 1 $\nu_{\tau}$ event, the significance of the result depends totally on 
the value of the
background. OPERA estimates the background to be $0.018\pm 0.007$ for the {\em 1-prong} decay channel where the candidate has been
observed. That corresponds to a probability of 1.8\% to fluctuate to 1 event, which
may be interpreted as a significance of 2.36 sigma's towards the observation of a $\nu_{\tau}$ interaction 
({\em p}-values of the null hypothesis, see Ref. [\refcite{PDG}]).

\begin{figure}[htb]
\begin{center}
\psfig{file=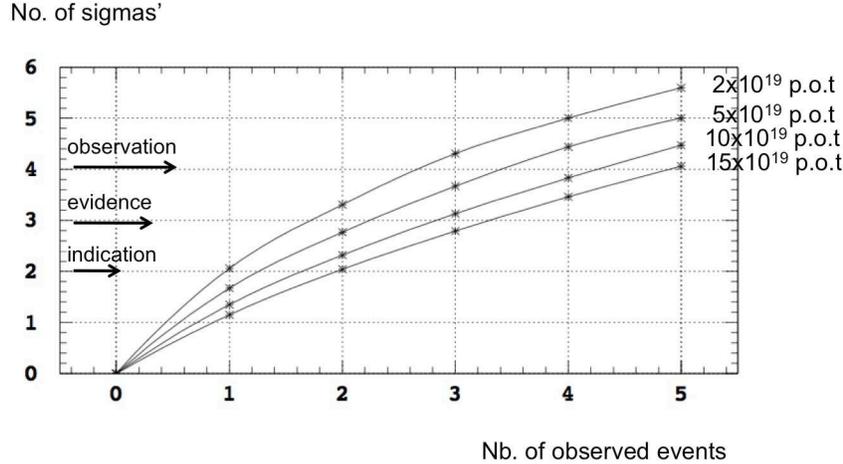,width=4.5in}
\end{center}
\caption{The number of Gaussian sigma's corresponding to the number of observed $\nu_{\tau}$ candidates. The 3 curves represent
different amount of data collections in terms of p.o.t. It also shown the level of confidence which is usually attributed to physics results
in terms of sigma's, as suggested by the author. The evaluation has been performed by considering the backgrounds from all the $\nu_{\tau}$
decay channels.}
\label{Prob1}
\end{figure}

At first sight it may be surprising to extract such level of significance from just one event. That is the power of a {\em clean} experiment.
It is ilustrated in Fig.~\ref{Prob1} where the significance of the result is drafted towards the number of events observed instead of the usual  
integrated 
{\em luminosity} of the data collected. The curves parametrized as function of the number of p.o.t collected by OPERA\footnote{The inputs
in terms of expectation of number of $\nu_{\tau}$ candidates and background events have been extracted and used from  the OPERA
proposal\cite{opera4} in 2001.}
show that very few events allow to set a quite robust physics result. On top of that it is also evident that whether OPERA will be able to
decrease the level of background the significance will increase it.
For example, in case the estimated background be increased/decreased of a factor 2, retaining the assumed 50\% nuisance, the corresponding
significances will decrease/increase as 2.10 and 2.61, respectively.
From another point of view the detection of a second (third) $\nu_{\tau}$ candidate, with the present level of total background proportionally updated,
will increase the statistical significance from 2.01 to 2.82 (3.42). The latter consideration may demonstrate that the OPERA result be potentially
much more interesting that the actual measurement by Super-Kamiokande which set a 2.4 significance in the $\nu_{\tau}$ appearance 
observation\cite{sk2}.

The OPERA result, at 98.2\% of probability, corresponds to an extremely important evidence which can be expressed in several ways. 
For example, we may say that it is the first direct evidence of Lepton Flavor Violation, the theoretical unsatisfaction of the Standard Models being
from now on even more evident. The observation of the transition from one flavor to the other should constraint and open new horizons to the 
theoretical elaborations, not forgetting the parallelism (somehow {\em opposite} in term of flavor eigenstates) with the quark sector.

\begin{figure}[htb]
\begin{center}
\psfig{file=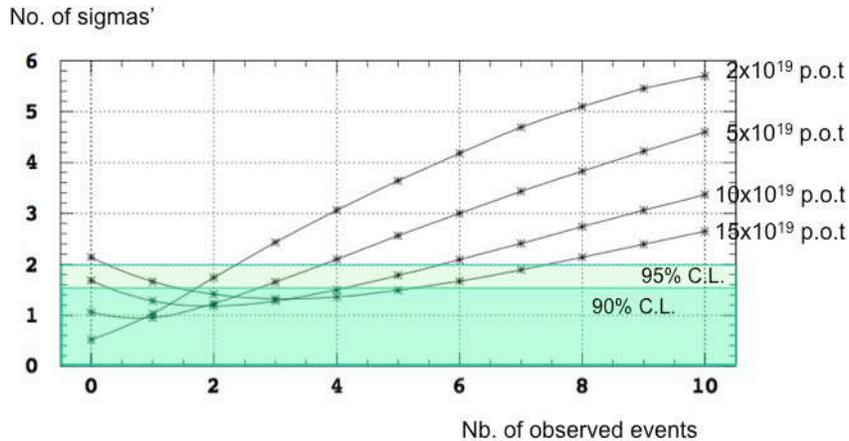,width=4.5in}
\end{center}
\caption{The {\em distance} in terms of number of Gaussian sigma's to the expectation from the MINOS result 
($\Delta m^2_{23} = 2.5\times 10^{-3}\, eV^2$) as a function of the observed $\nu_{\tau}$ candidates. The 3 curves represent
different amount of data collections in terms of p.o.t. The two level of confidence, at 90\% and 95\% are also shown. 
The detection of ZERO candidates is marginal even after $15\times 10^{19}$ p.o.t. analyzed (but affordable), while departure from the expectation
is possible even with very few candidate events.}
\label{Prob2}
\end{figure}

The second important point which is left to OPERA for the near future is to answer the question about the number of oscillated $\nu_{\tau}$.
That issue is well illustrated by a plot similar to the previous one (Fig.~\ref{Prob2}) where the {\em distance} in terms of sigma's from
the MINOS expectation is drawn towards the number of observed events. The result is parametrized as function of number of p.o.t.
From the figure we may deduce that it will take some time to disentangle any deviation from the standard oscillation scenario.
However it is will be fully worthwhile to pursue it.

\section{Conclusions}

The neutrino oscillation scenario began to be clarified in 1998 with the observation of a disappearance of atmosferic $\nu_{\mu}$, followed by
the determination of similar disappearance (and indirect appearance) in the solar sector.
The scenario that rose up is based on a 3-flavor oscillation which however leave out some intriguing concerns like the LSND result and the
presence or not of {\em sterile} neutrinos. In that context possible correlations with the similar mixing pattern of the quark sector 
are still at the level of theoretical exercises. The powerful results by MINOS settled a stringent measurement on the $\nu_{\mu}$ oscillation.
The very recent result by OPERA, even if still at the level of {\em evidence}, demonstrates the action of LFV and it rules out for the time
being the presence of {\em sterile} neutrinos. 
The large numbers of experiments undergoing all over the world to search for a $\theta_{13}$ value different from zero corresponds  to a lively
field of physics interest (see other contributions on these proceedings). However more than usual it is necessary to outline the
lesson from past, nature is not obvious and the lack of experimental confirmations about theoretical models should encourage us
to be prepared on the unexpected.

\section{Acknowledgments}

It is a pleasure to thanks the very warm hospitality of the organizers and the kind invitation.
The presentation allowed the author to further elaborate on the very attractive field of neutrino physics.
Some results and discussions reported in these proceedings were stimulated just for 
this occasion. Some of the statistical elaborations have been checked through by my colleague S. Dusini.
Also I want to thank M. Mezzetto for a critical reading of the draft, many of his suggestions have been 
incorporated in the present version. Finally, all the considerations and the conclusions throughout the paper
are of full responsibility of the author.

\bibliographystyle{ws-procs975x65}
%\bibliography{ws-pro-sample}

\end{document}